\documentclass[twocolumn,nofootinbib]{revtex4}
\usepackage{multirow}
\usepackage{pstricks}
\usepackage{dcolumn}
\usepackage{bm}
\usepackage{latexsym}
\usepackage{dcolumn}
\usepackage{amsmath}
\usepackage{amsfonts,amssymb}
\usepackage{graphicx,epsfig}
\usepackage{color}
\usepackage{psfrag}
\usepackage{amsthm}
\usepackage{ulem} 
\usepackage{soul} 
\newcommand\Tstrut{\rule{0pt}{2.6ex}}         
\newcommand\Bstrut{\rule[-0.9ex]{0pt}{0pt}}   

\def\be{\begin{equation}}
\def\ee{\end{equation}}
\def\be{\begin{equation}}
\def\ee{\end{equation}}
\def\bea{\begin{eqnarray}}
\def\eea{\end{eqnarray}}

\def\bes{\begin{equation*}}
\def\ees{\end{equation*}}
\usepackage{slashed}
\def\met{\slashed E_T}
\def\lsim{\raise0.3ex\hbox{$\;<$\kern-0.75em\raise-1.1ex\hbox{$\sim\;$}}}
\def\gsim{\raise0.3ex\hbox{$\;>$\kern-0.75em\raise-1.1ex\hbox{$\sim\;$}}}

\begin{document}
\title{Dark Matter Spin  Characterisation in Mono-$Z$ Channels}
\author{W. Abdallah$^{1,2}$, A. Hammad$^{3}$, S. Khalil$^{4}$ and S. Moretti$^{5}$}
\affiliation{
$^1$Harish-Chandra Research Institute, Chhatnag Road, Jhunsi, Allahabad 211019, India\\
$^2$Department of Mathematics, Faculty of Science, Cairo University, Giza 12613, Egypt\\
$^3$Department of Physics, University of Basel, Klingelbergstra\ss e 82, CH-4056 Basel, Switzerland\\
$^4$Center for Fundamental Physics, Zewail City of Science and Technology, 6 October City, Giza 12588, Egypt\\
$^5$School of Physics and Astronomy, University of Southampton, Highfield, Southampton SO17 1BJ, UK}
\begin{abstract}
\noindent
The $B-L$ Supersymmetric Standard Model (BLSSM) is an ideal testing ground of the spin nature of Dark Matter (DM) as it offers amongst its candidates both a spin-1/2 (the lightest neutralino) and spin-0 (the lightest right-handed sneutrino) state.
We show that the mono-$Z$ channel can be used at the Large Hadron Collider (LHC) to diagnose whether a DM signal is characterised within the BLSSM by a fermionic or (pseudo)scalar DM particle. Sensitivity to either hypothesis can be obtained after only 100~fb$^{-1}$ of luminosity following Runs 2 and 3 of the LHC.    
\end{abstract}
\maketitle
\section{Introduction}

DM is one of the firm evidences of physics Beyond the Standard Model (BSM). Searches for DM at the  LHC  through Missing Transverse Energy (MET or $\met$) and probing a single particle, like mono-jet, -photon, -$Z$ and -Higgs, are one of the most promising methods for establishing DM  existence directly in an experiment. However, the nature of  DM remains as one of the foremost open questions in particle physics, especially whether the DM is a fermionic or bosonic particle. 

Fermionic DM is predicted by several BSMs, like the Minimal Supersymmetric Standard Model (MSSM), in which the lightest neutralino (a fermionic superpartner of the neutral scalar and gauge bosons of the SM) is a quite popular example of weak scale DM. Scalar DM has been analysed in models with extra inert singlet or doublet Higgs bosons.  Here, we will perform a comparative study for the two types of DM, predicted by the same model, the BLSSM, in different regions of  parameter space. 

The BLSSM is a natural extension of the MSSM with an extra $U(1)_{B-L}$. It accounts for non-vanishing neutrino masses through a low scale seesaw mechanism, which can be an inverse seesaw (see Ref.~\cite{Khalil:2015naa} for a review). In this scenario, it is quite possible to have the lightest neutralino or the lightest right-handed sneutrino as the Lightest Supersymmetric Particle (LSP), so that any of these can be a stable DM candidate~\cite{Abdallah:2017gde}. A detailed analysis of BLSSM DM candidates has been performed in~\cite{Abdallah:2017gde,DelleRose:2017uas} (see also~\cite{DelleRose:2017ukx}). Therein, it was shown that, for a wide region of parameter space, the lightest right-handed sneutrino, with mass of order ${\cal O}(100)$~GeV, can be a viable DM candidate that satisfies the limits of relic abundance and also the scattering cross sections with nuclei. The chances of the lightest neutralino being the actual DM state are much less in comparison, however, in some regions of the parameter space, it is still possible to have it as the origin of DM, in particular,  in the form of the  lightest $B-L$ neutralino. Further, in Ref.~\cite{DelleRose:2017uas}, it was shown that the Fermi Large Area Telescope (FermiLAT) can be sensitive to the DM spin (and eventually distinguish between the sneutrino and neutralino hypotheses) in the study of high-energy $\gamma$-ray spectra emitted from DM (co)annihilation  into $W^\pm$ boson pairs (in turn emitting photons). 

Furthermore, we studied several single-particle signatures of the BLSSM DM at the LHC, i.e., 
mono-jet, -photon, -$Z$ and -Higgs signals, 
induced by new channels mediated by the heavy $Z'$ (in the few~TeV range) pertaining to the (broken) $U(1)_{B-L}$ group
\cite{Abdallah:2015uba,Abdallah:2016vcn}.  The salient feature of this BLSSM specific channel is that the final state mono-probe carries a very large MET. Hence, it is a clean signal,  almost free from  SM background.  It was argued that, with luminosities of order $100~{\rm fb}^{-1}$, mono-jet events associated with BLSSM DM can be accessible at the LHC while mono-photon, -$Z$  and -Higgs signals can be used as diagnostic tools of the underlying scenario.

In this paper, we expand on all these results, by 
showing that DM spin can be accessed at the LHC in the mono-$Z$ channel. We prove  this result by showing that the angular distributions of the final state lepton emerging from a subsequent $Z$ decay, for both  neutralino and  right-handed sneutrino DM, are significantly different from each other. This is in contrast to the result that  these distributions are identical in mono-jet, -photon and -Higgs (owing to the fact that jets and $\gamma$'s do not couple 
directly to DM while Higgs radiation is isotropic), thus being insensitive to the DM spin.

This paper is organised as follows. In Sect.~\ref{sec.2} we briefly highlight the possibility of having both (pseudo)scalar and  fermionic DM in the BLSSM with {an}  inverse seesaw mechanism. Sect.~\ref{sec.3} is dedicated to the mono-$Z$ analysis in these two DM scenarios. In Sect.~\ref{sec.4} we discuss the impact of the DM spin on the angular distributions of the corresponding final leptons. Our conclusions and final remarks are given in Sect.~\ref{sec.5}.

\section{Scalar versus Fermionic DM}
\label{sec.2}
The BLSSM is based on the gauge group $SU(3)_C\times SU(2)_L\times U(1)_Y\times
U(1)_{B-L}$, where the $U(1)_{B-L}$ is spontaneously broken at the TeV scale~\cite{Khalil:2016lgy} by 
chiral singlet superfields $\hat{\eta}_{1,2}$ with $B-L$ charge $=\pm 1$.
Here, a gauge boson $Z'$ and three chiral singlet
superfields $\hat{\nu}_i$ with $B-L$ charge $=-1$ are introduced for
the consistency of the model. Finally, three chiral singlet
superfields $\hat{S}_1$ with $B-L$ charge $=+2$ and three chiral singlet
superfields $\hat{S}_2$ with $B-L$ charge $=-2$ are considered to
implement the inverse seesaw 
mechanism~\cite{Khalil:2010iu}. The superpotential  is given by%
\bea
W &=&  Y_u\hat{Q}\hat{H}_2\hat{U}^c + Y_d \hat{Q}\hat{H}_1\hat{D}^c+ Y_e\hat{L}\hat{H}_1\hat{E}^c\nonumber\\
&+&Y_\nu\hat{L}\hat{H}_2\hat{\nu}^c+Y_S\hat{\nu}^c\hat{\eta}_1\hat{S}_2 +\mu\hat{H}_1\hat{H}_2+ \mu'\hat{\eta}_1\hat{\eta}_2. 
\label{superpotential}
\eea
The neutralinos, $\tilde{\chi}^0_i $ ($i=1,\dots,7$), are
the physical (mass) superpositions of the three fermionic partners of the 
neutral gauge bosons, called gauginos, 
of the neutral MSSM Higgs bosons ($\tilde{H}_1^0$ and $\tilde{H}_2^0 $), called Higgsinos, and of the $B-L$ scalar bosons ($\tilde{\eta}_1$ and $\tilde{\eta}_2 $). In
this regard, the lightest neutralino, in the basis $\psi^0=\{{\tilde B},{\tilde
W}^3,{\tilde H}^0_1,{\tilde
H}^0_2,{\tilde B'},{\tilde \eta_1},{\tilde
\eta_2}\}$, decomposes as  
\be 
\tilde\chi^0_1=\sum_{i=1}^7V_{1i}\psi^0_i. 
\ee 
The lightest sneutrino $\tilde{\nu}_1$ (either a CP-even state, $\tilde{\nu}^{\rm R}_1$, or a CP-odd one, $\tilde{\nu}^{\rm I}_1$) can be expressed in terms of $\tilde{\nu}^+_L$, $\tilde{\nu}^+_R$ {and} $\tilde{S}^+_2$ (e.g., in case of it being CP-even) as 
\be 
\tilde{\nu}_1 = \sum_{i=1}^3 R_{1i} (\tilde{\nu}^+_L)_i +  \sum_{j=1}^3 R_{1j} (\tilde{\nu}^+_R)_j +  \sum_{k=1}^3 R_{1k} (\tilde{S}^+_2)_k,
\label{Gamma}
\ee
where $R_{1i} \approx  \{ 0,0,0\} $, $R_{1j} =\frac{1}{\sqrt{2}} \{ 1,0,0\}$, and $R_{1k} =\frac{1}{\sqrt{2}} \{ 1,0,0\}$, which confirms that the lightest sneutrino is mainly right-handed ({i.e.}, a combination of $\tilde{\nu}^+_R$ and $\tilde{S}^+_2$). 

\begin{figure}[h!]
\includegraphics[width=10cm,height=6cm]{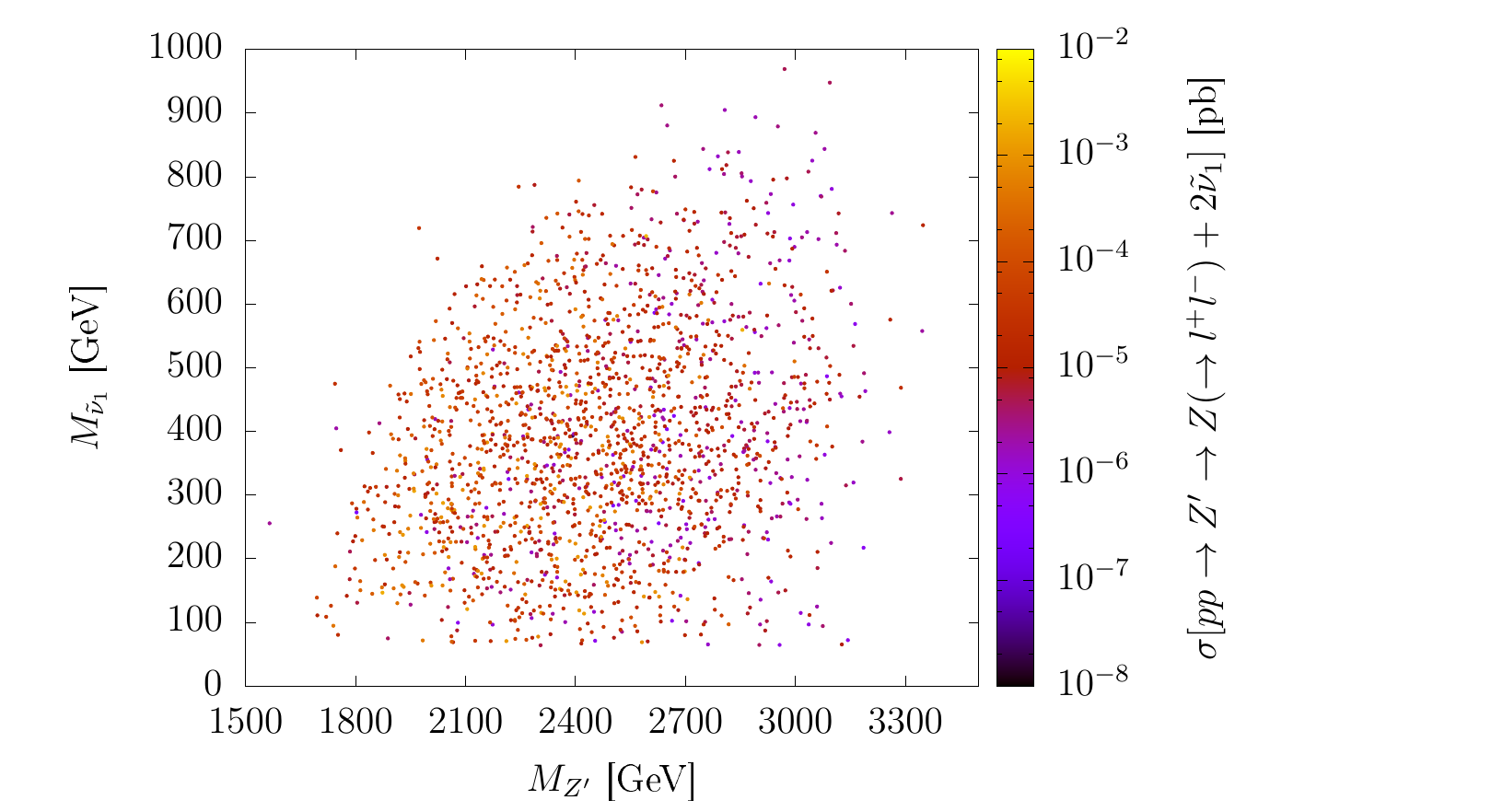}
\caption{The cross section for $pp\to Z'\to Z(\to l^+l^-)  + 2 \tilde\nu_1$ at the LHC with $\sqrt s=$ 14~TeV mapped over the $Z'$ and $\tilde\nu_1$ masses for the BLSSM with {an} inverse seesaw {mechanism}. 
}
\label{MZp-xsec-Msv1}
\end{figure}

\begin{figure*}[t!]
\centering
\includegraphics[width=6.cm,height=3.5cm]{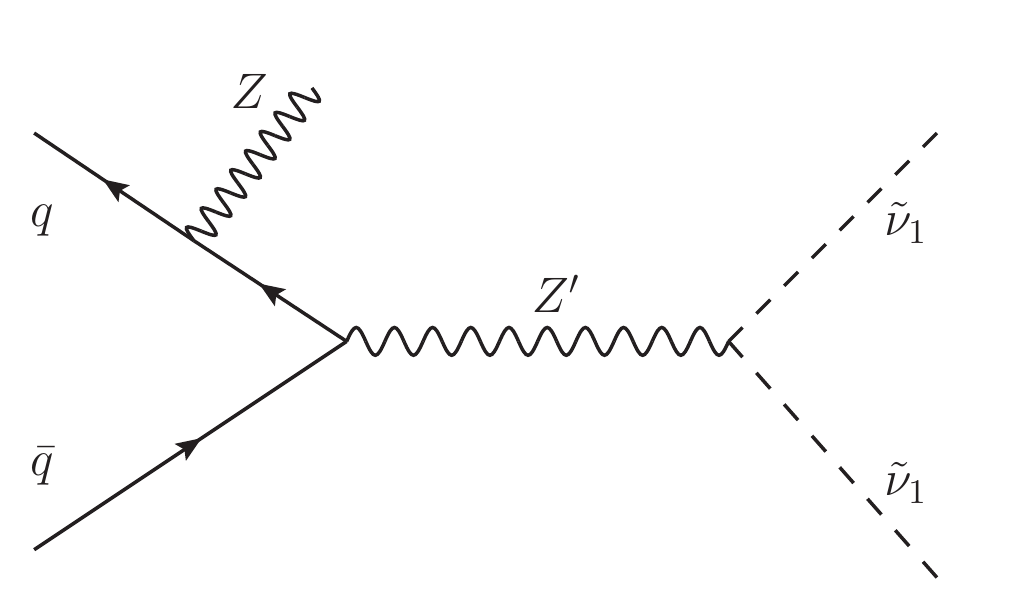}~~~~~~~
\includegraphics[width=6.cm,height=3.5cm]{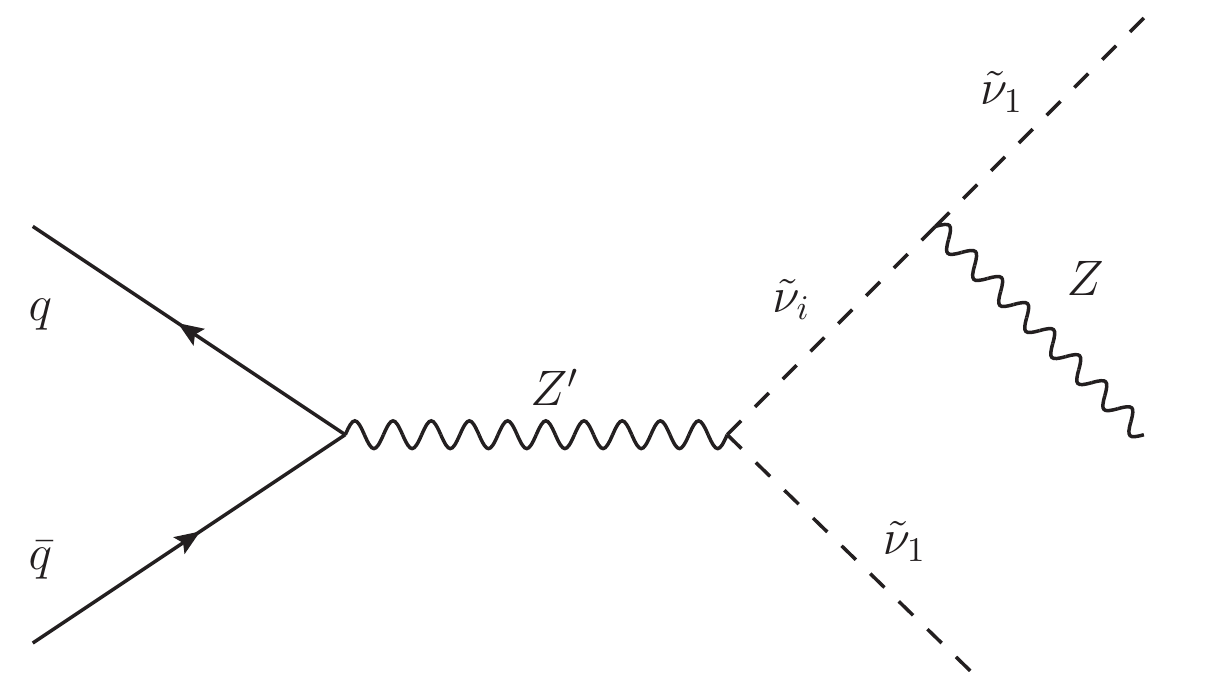}\\[0.2cm]
\includegraphics[width=6.cm,height=3.5cm]{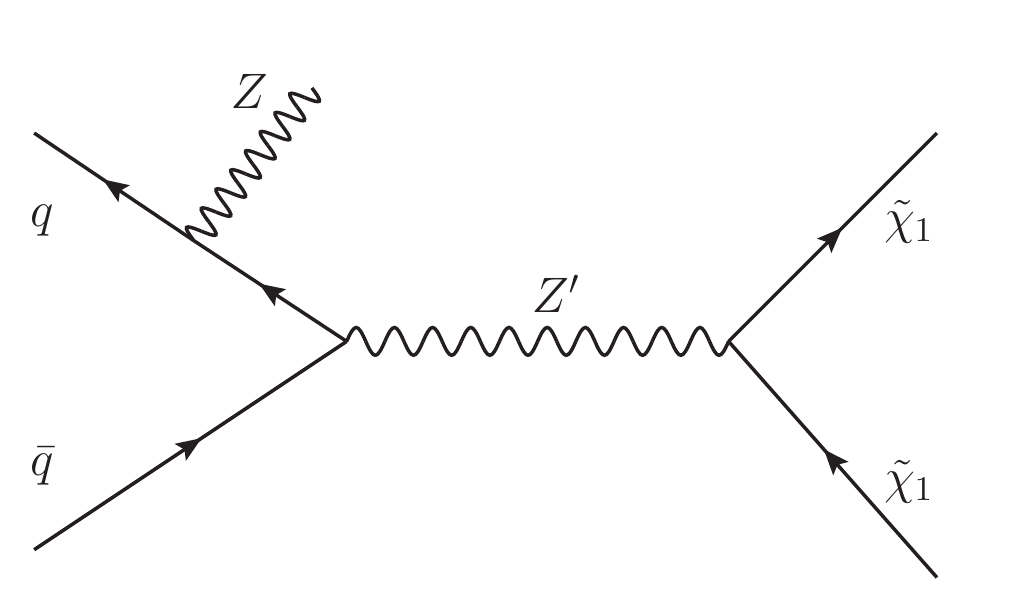}~~~~~~~
\includegraphics[width=6.cm,height=3.5cm]{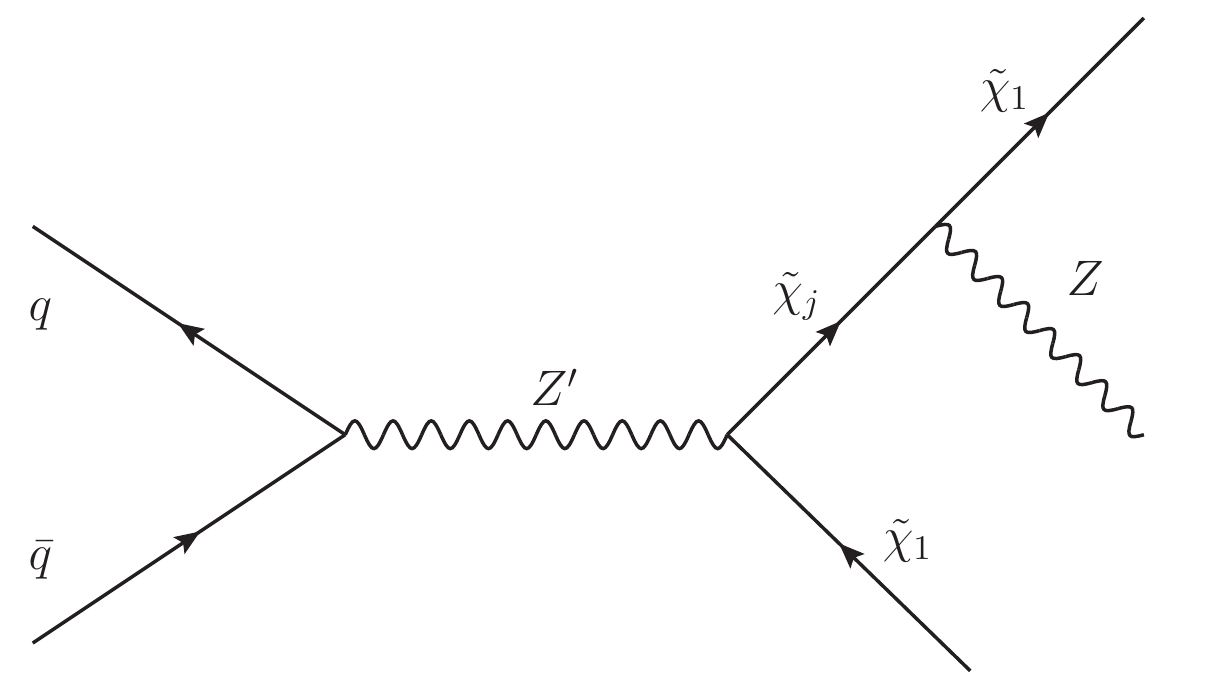}
\caption{Feynman diagrams for mono-$Z$ signals {of scalar (top panels) or fermionic (bottom panels) DM}: $S_1$ (left) and $S_2$ (right), corresponding to $Z$ Initial State Radiation (ISR) and Final State Radiation (FSR), respectively. {Here, $i=2,3,\dots,9$ and $j=2,3,\dots,7$.}}
\label{fyn}
\end{figure*}

It is worth mentioning that, due to the $U(1)_Y$ and $U(1)_{B-L}$ gauge kinetic mixing, the mass of the extra neutral gauge boson, $Z'$, is given by
\be
M_{Z'}^2 = g_{_{B-L}}^2 v'^2 + \frac{1}{4} \tilde{g}^2 v^2 ,
\ee
where $\tilde{g}$ is {the gauge kinetic mixing} coupling. Also, the mixing angle between $Z$ and $Z'$, which is experimentally limited to $\lsim {{\cal O}}(10^{-3})$, is given by
\be
\tan 2 \theta' = \frac{2 \tilde{g}\sqrt{g_1^2+g_2^2}}{\tilde{g}^2 + {4} (\frac{v'}{v})^2 g_{_{B-L}}^2 -g_2^2 -g_1^2}.
\ee
The relevant interactions of the lightest neutralino and  lightest right-handed sneutrino with the $Z'$ and $Z$ bosons are given by 
\small{
\bea
\hspace{-1cm}&{\cal L}_{{\rm int}}&  \simeq - i\left(\frac{\tilde{g}}{2}\Delta V_{34}+g_{_{B-L}}\Delta V_{67}\right)\overline{\tilde{\chi}_1^0}\slashed{Z'}\gamma_5\tilde{\chi}_i^0\nonumber\\
\hspace{-1cm}&+&\!\!\!\!\!\!\!\frac{g_{_{B-L}}}{2}\displaystyle\sum_{n=1}^3\tilde\nu^{\rm R}_1\tilde\nu^{\rm I}_j (p'-p)^\mu Z'_\mu \left(I^*_{j,n+6} R^*_{1,n+6}-I^*_{j,n+3} R^*_{1,n+3}\right)\nonumber\\
\hspace{-1cm}&-&\!\!\!\!\!\!\frac{i}{2}\big[(g_2 \cos {\theta_W}+g_1 \sin {\theta_W})\Delta V_{34}-2\tilde{g}\Delta V_{67}\big]\overline{\tilde{\chi}_1^0}\slashed{Z}\gamma_5\tilde{\chi}_i^0\nonumber\\
&+&\!\!\!\!\!\!\frac{1}{2}(g_2 \cos {\theta_W}+g_1 \sin {\theta_W})\displaystyle\sum_{n=1}^3\tilde\nu^{\rm R}_1\tilde\nu^{\rm I}_j (p'-p)^\mu Z_\mu I^*_{j,n} R^*_{1,n},
\eea 
}
where $\Delta V_{nm}=V^*_{in}V_{1n}-V^*_{im}V_{1m}$. Fig.~\ref{MZp-xsec-Msv1} shows the total cross section for $pp\to Z^\prime \to  Z(\to {l^+}l^-) +2 \tilde{\nu}_1$ ($l=e,\mu$),
based on the diagrams {(top panels)} in Fig.~\ref{fyn} (summed and  squared, thereby capturing the relative interference too), for different masses  of the $Z^\prime$ and $\tilde{\nu}_1$ after satisfying all Higgs data constraints by using HiggsBounds~\cite{Bechtle:2008jh} and HiggsSignals~\cite{Bechtle:2013xfa}. The scanned points have been generated over the following intervals of the BLSSM fundamental parameters:  $ 10^3$~TeV $ \le M^2_{\tilde{l}}\le 5\times 10^3$~TeV, $ -500$~TeV $ \le M^2_{\tilde{\nu}}\le -10^2$~TeV, $ 10$~TeV $ \le M^2_{\tilde{S}}\le 50$~TeV, $ 0.3 \le g_{_{B-L}} \le 0.5$, $ -0.4 \le \tilde{g} \le -0.2$, $4$~TeV $ \le v'_1 \le 6$~TeV {and} $3$~TeV $ \le v'_2 \le 5$~TeV  plus, to ensure that the lightest $\tilde{\nu}_1$ is the LSP, we kept $ M_1 = M_2 = M_3 = 6$~TeV. A benchmark point will be  chosen from the scanned ones to perform a detailed Monte Carlo analysis. As the latter will be based around $Z'$ production and decay, we also have made sure that, on the one hand, the scan points do not fall out of the LEP (indirect) constraints and, on the other hand, the ensuing $Z^\prime$ will not have been discovered via LHC (direct) searches in Drell-Yan (DY) mode already.  We meet these conditions by adjusting the parameters of the chosen point as follows:   $M_{Z^\prime} = 2.9$~TeV, $M_{\tilde{\nu}_1}\simeq 90$~GeV, $g_{_{B-L}} = 0.5$ and $\tilde{g}= -0.25$. 
\vspace{-0.03cm}
\section{Mono-$Z$ analysis}
\label{sec.3}
In the following, we will develop an analysis aimed at extracting information about the {lightest right-handed} sneutrino of the BLSSM as the DM candidate through a dedicated mono-$Z$ search using a Machine Learning (ML) algorithm called Boosted Decision Tree (BDT)~\cite{Alves:2015dya,Alves:2017uls}. The key to this approach is to rely on a mono-jet evidence of DM in a kinematic regime compatible with $Z'$ production and decay\footnote{{{Contrary to Ref.~\cite{Bernreuther:2018nat}, here, the contributions of the $Z$ and SM-like Higgs ($h_{\rm SM}$) as mediators are very small due to a lower bound on the LSP mass, in particular, one has $M_{\tilde{\nu}_1} > M_{h_{\rm SM}}/2> M_Z/2$, in order to satisfy the invisible SM-like Higgs decay upper limit~\cite{Aaboud:2018sfi,Sirunyan:2018owy}. This means that $Z$ and $h_{\rm SM}$ propagators are off-shell, unlike the $Z'$ one. Further, the $Z'$   couplings to sneutrinos aremuch  stronger than those of the $Z$ and $h_{\rm SM}$. Finally, we will enforce a stiff MET cut to enhance the $Z'$ component of the signal.}}}, so that, under a model dependent assumption (i.e., assuming the BLSSM), one can extract mono-$Z$ signatures leading to the identification of the DM properties, chiefly, of its spin.
In fact, an intriguing feature of the mono-$Z$ analysis is the possible spin characterisation  of DM. Spin determination methods rely heavily on the final state spins and the chiral structure of the couplings. The 2-body decays of neutralinos to a massive $Z$ boson and a DM neutralino produce a $Z$ boson in three helicity states, $\pm 1$ (transverse) and $0$ (longitudinal). Reconstructing the three polarisation states through the angular distributions of the $Z$ boson leptonic decays through $\tilde{\chi}_i^0\to \tilde{\chi}_1^0 Z(\to l^+ l^-)$  in the rest frame of the decaying $Z$ boson leads to a clear characterisation of the spin state  of the $Z$ boson. The angular distribution  of the transverse states are $\propto (1\pm \cos^2\theta)$ while the angular distribution of the longitudinal state is $\propto \sin^2\theta$, where $\theta$ is the angle between the lepton momentum {direction} and the $Z$ boson one in the latter rest frame. The decay width of the neutralino $\tilde{\chi}_i^0$ to Transversely ($T$) and {L}ongitudinally ($L$) polarised $Z$ bosons is given by~\cite{Choi:2003fs}. It is worth mentioning that the decay width of the  longitudinal component of a $Z$ boson is suppressed with respect to its transverse ones~\cite{Choi:2018sqc}.

The 2-body decays of {heavier sneutrinos} to a massive $Z$ boson and sneutrino DM, $\tilde{\nu}_i\to \tilde{\nu}_1 Z(\to l^+ l^-)$,  produce a $Z$ boson in a zero-helicity (longitudinal) state only. This is because the helicity has to be conserved in the $S$-matrix and the fact that $\tilde{\nu}_i$ and $\tilde{\nu}_1$  are (pseudo)scalars forces the produced $Z$ boson to have a unique  state {(\it{cf}} Fig.~1 in~\cite{Choi:2018sqc}).
\begin{figure}[t!]
\includegraphics[width=7cm,height=5cm]{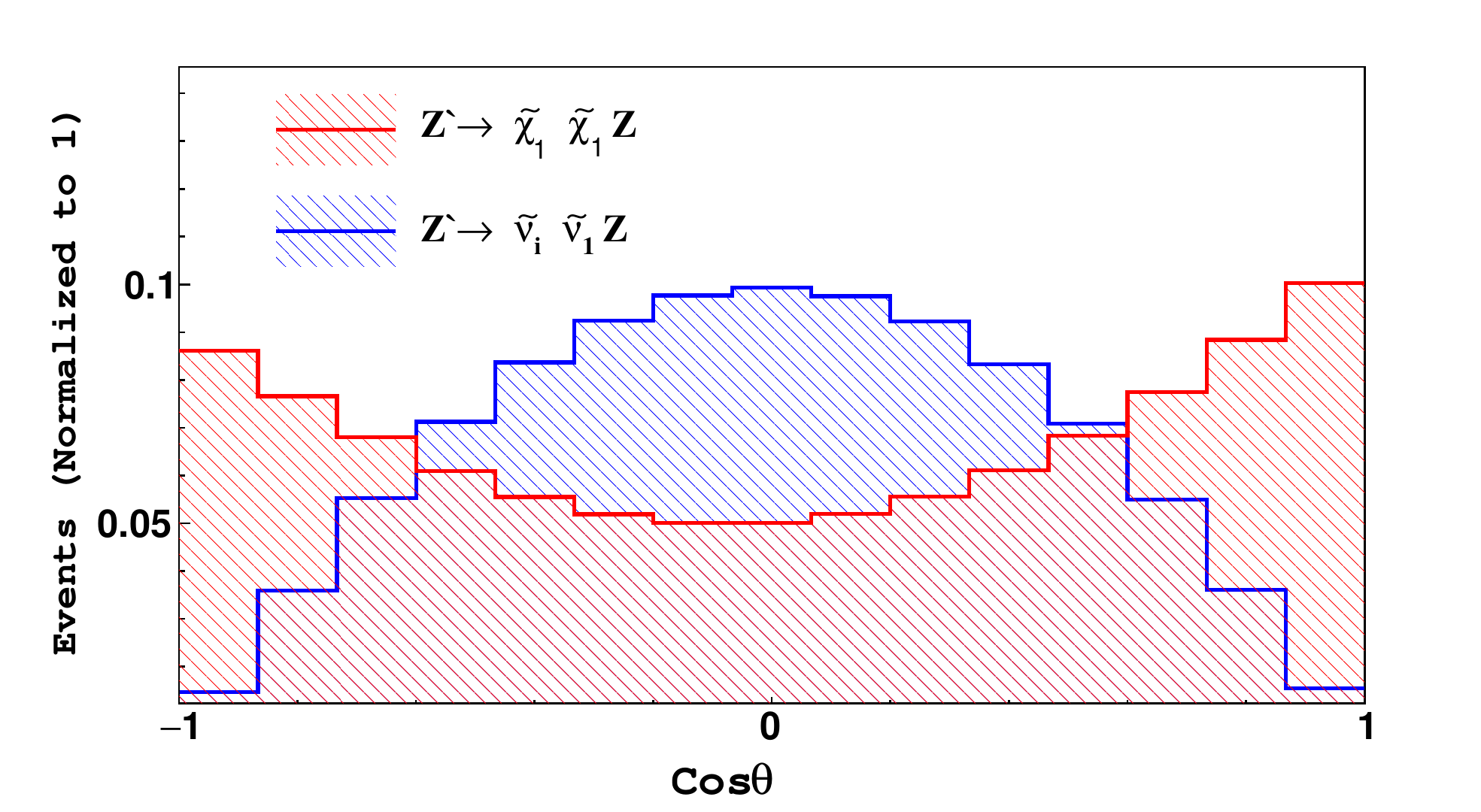}
\caption{Angular distribution of the final state lepton ($l = e, \mu)$ in presence of a neutralino mediator in red (transverse polarisation) and sneutrino mediator in blue (longitudinal polarisation), where $\theta$ is the angle between the lepton and $Z$ boson directions in the $Z$ rest frame.}
\label{polarisation}
\end{figure}
Fig.~\ref{polarisation} shows the angular distribution of the final state lepton $l$ for $\tilde{\chi}_i^0\to\tilde{\chi}_1^0 Z$ transitions in red and that for $\tilde{\nu}_i\to \tilde{\nu}_1 Z$ ones  in blue.  
It is also worth noting that, in Refs.~\cite{Dutta:2019gox,Bagger:1996bt,Ghosal:1997dv}, a similar approach based on angular distributions of leptonic $Z$ boson decays emerging from $\tilde{\chi}_1^0\to Z \tilde{G}$ transitions, with $\tilde G$ being light gravitino, was considered to distinguish between a  Higgsino- and gaugino-like neutralino in a model with Gauge-Mediated Supersymmetry Breaking (GMSB).

\section{Results}
\label{sec.4}
\begin{figure*}[t!]
\centering
\includegraphics[width=8.cm,height=5cm]{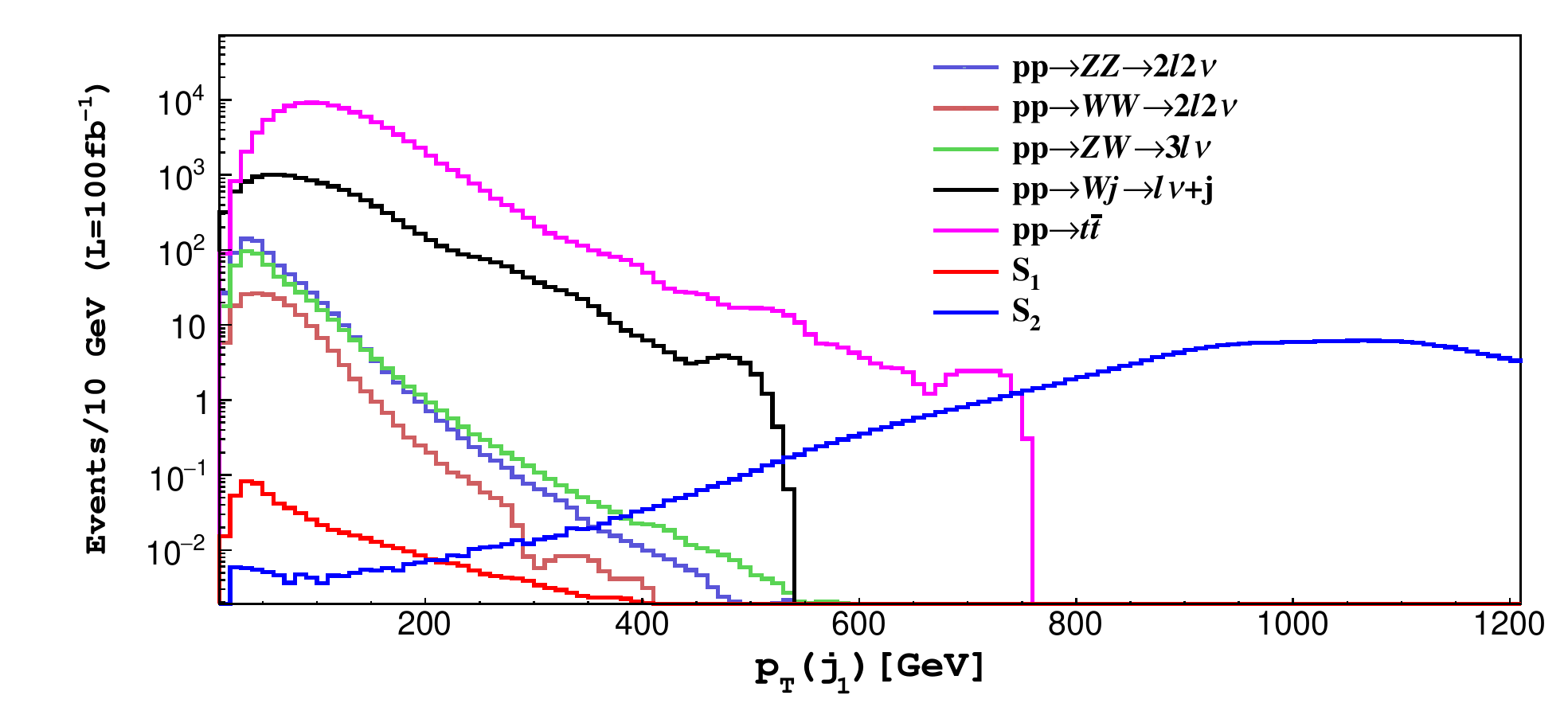}~
\includegraphics[width=8.cm,height=5cm]{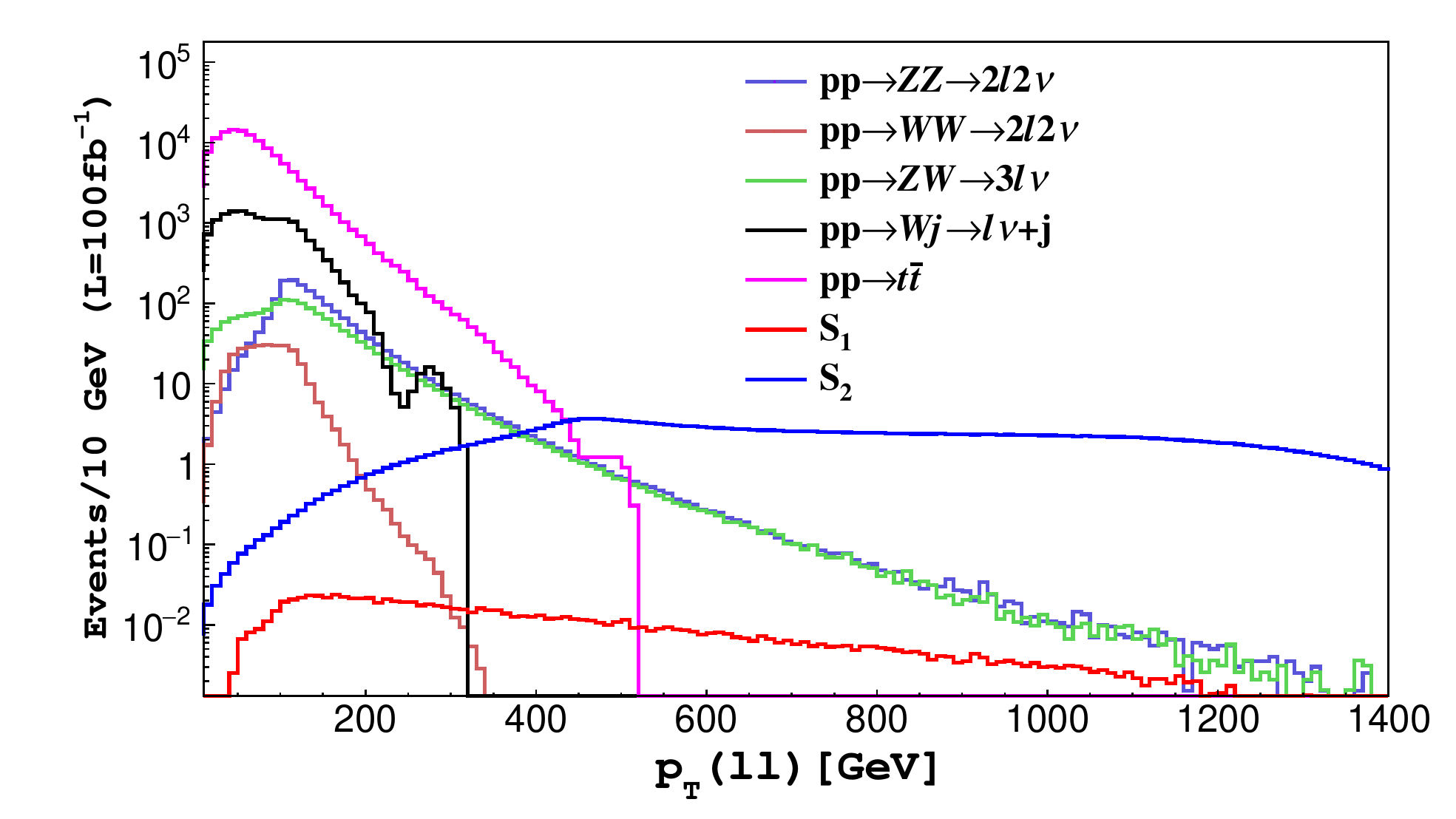}~
\caption{Transverse momentum of the leading jet (left) and of the di-lepton final state (right), with $S_1$ the signal process with $Z$ ISR (Fig.~\ref{fyn} left) and $S_2$ the signal process with $Z$ FSR  (Fig.~\ref{fyn} right).}
\label{Mll}
\end{figure*}

\begin{figure*}[t!]
\centering
~~~~\includegraphics[width=5.cm,height=4cm]{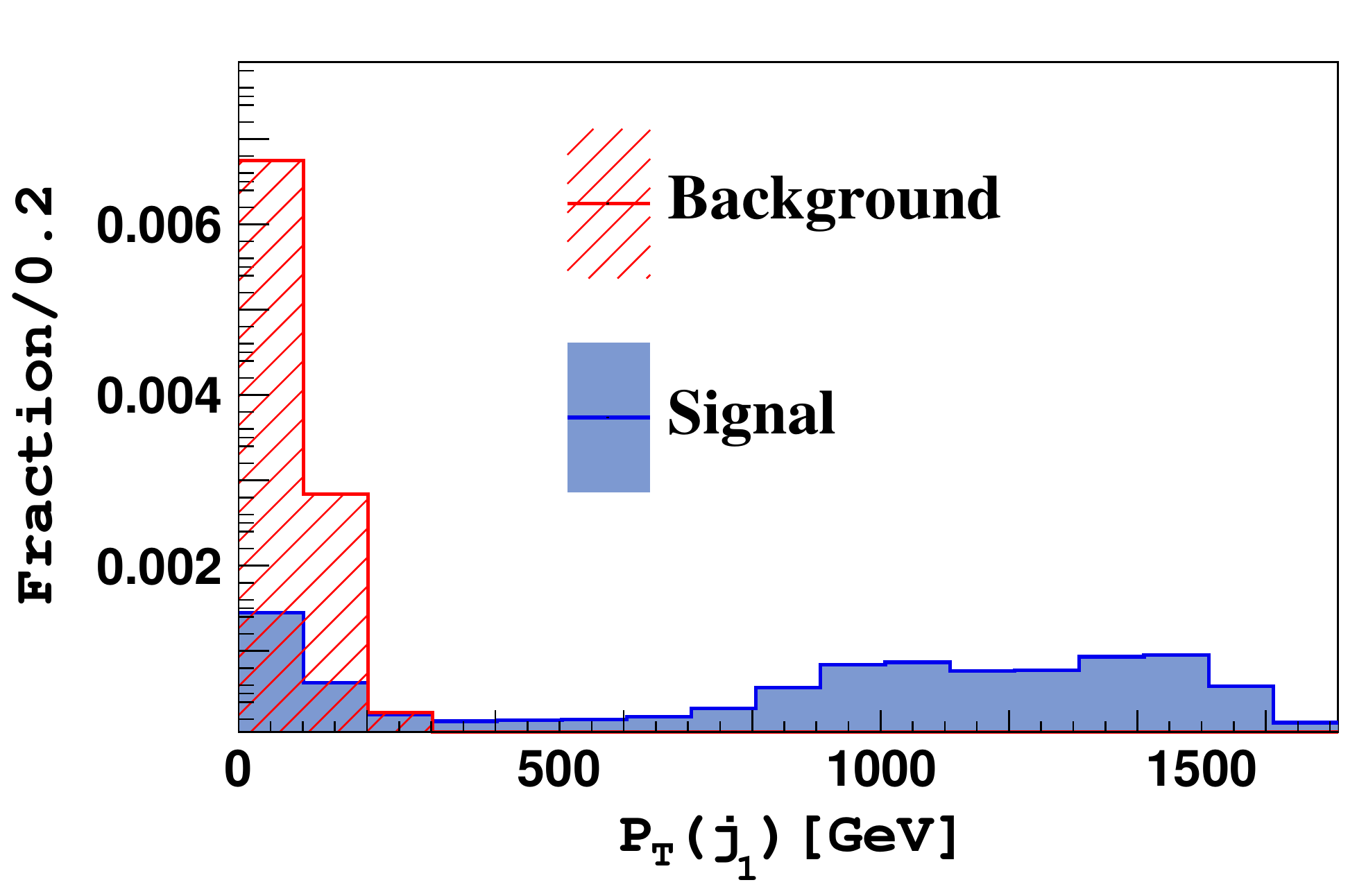}~~~\includegraphics[width=5.cm,height=4.4cm]{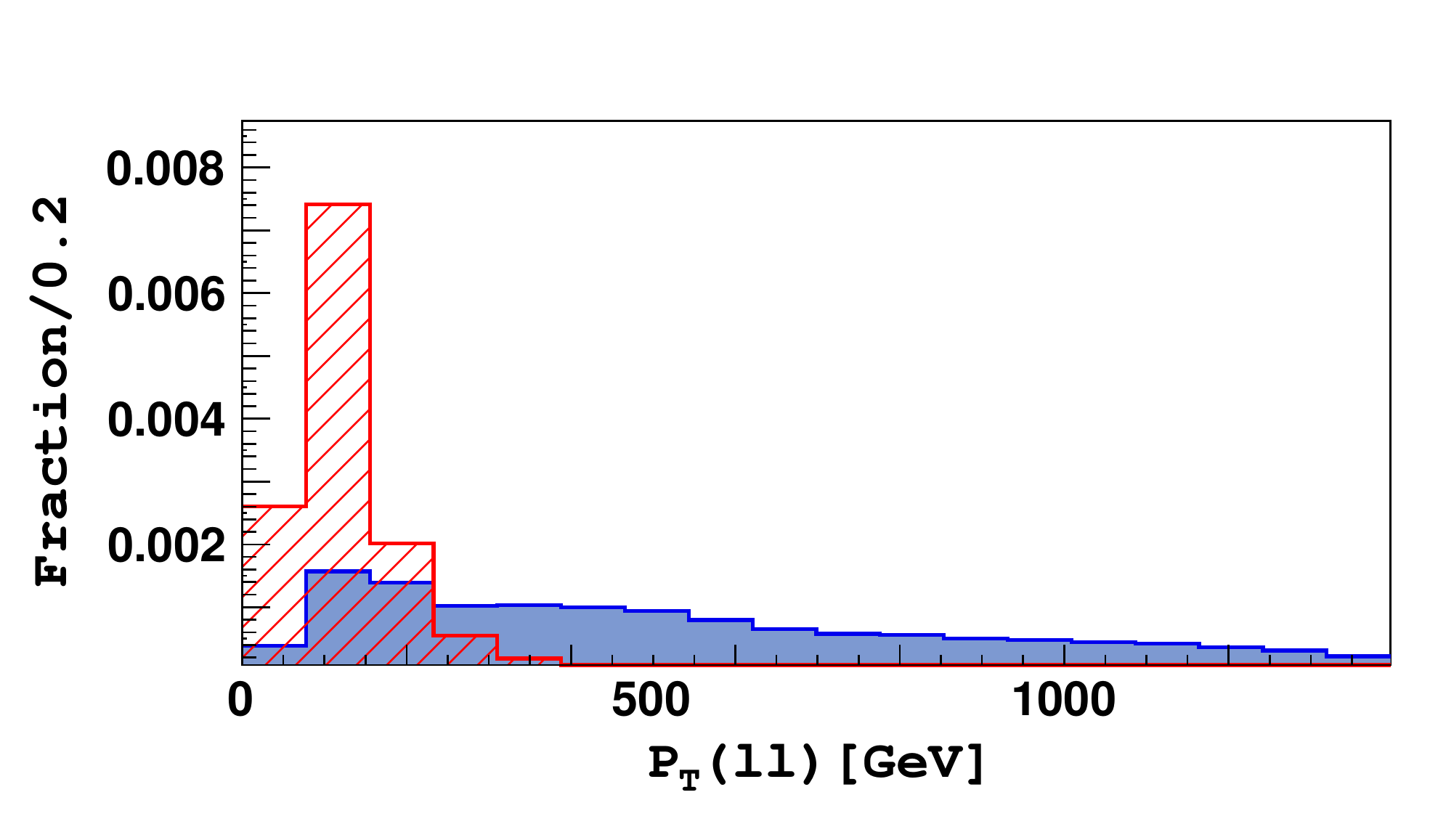}~~~\includegraphics[width=5.cm,height=4cm]{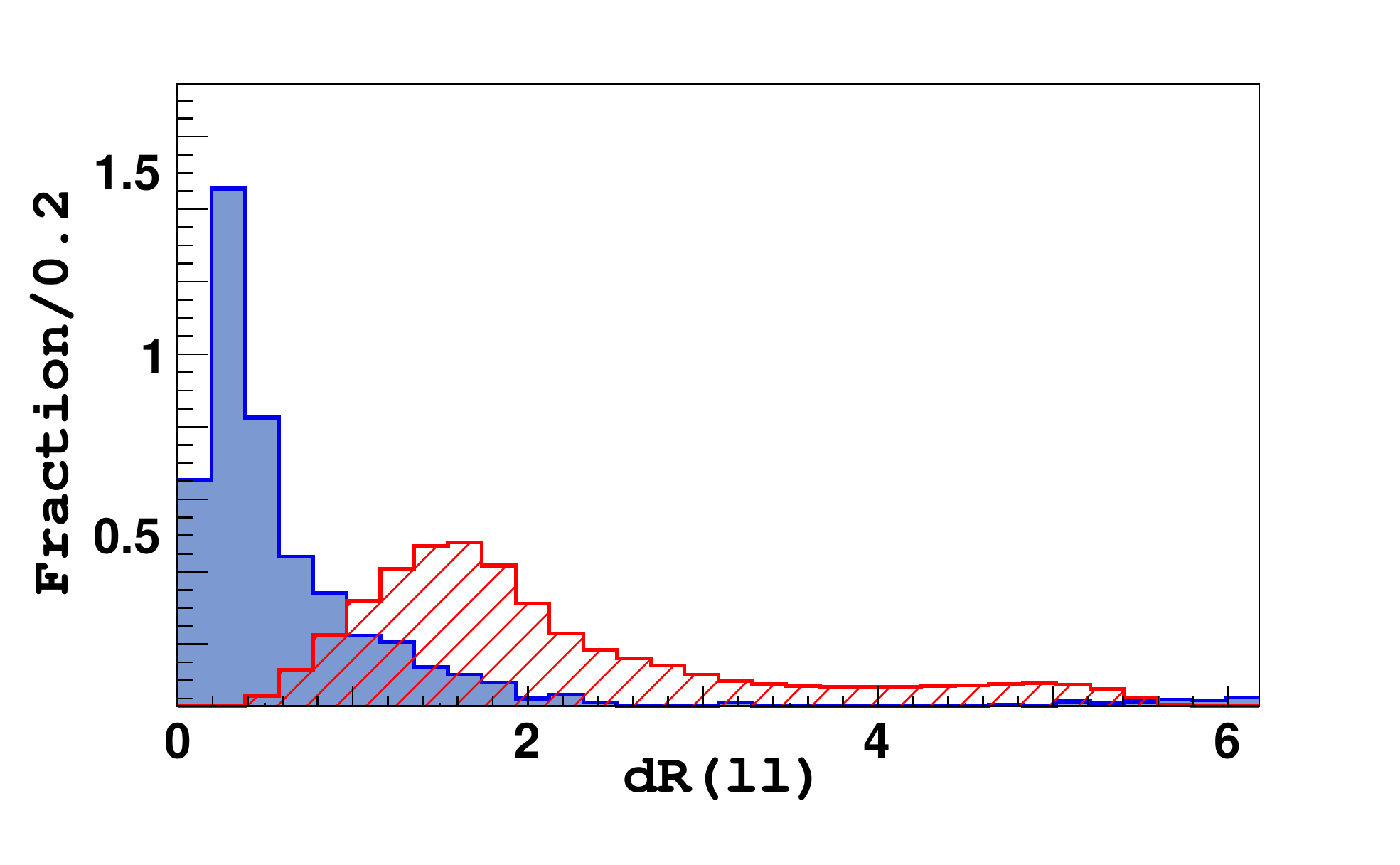}
\includegraphics[width=5cm,height=3.5cm]{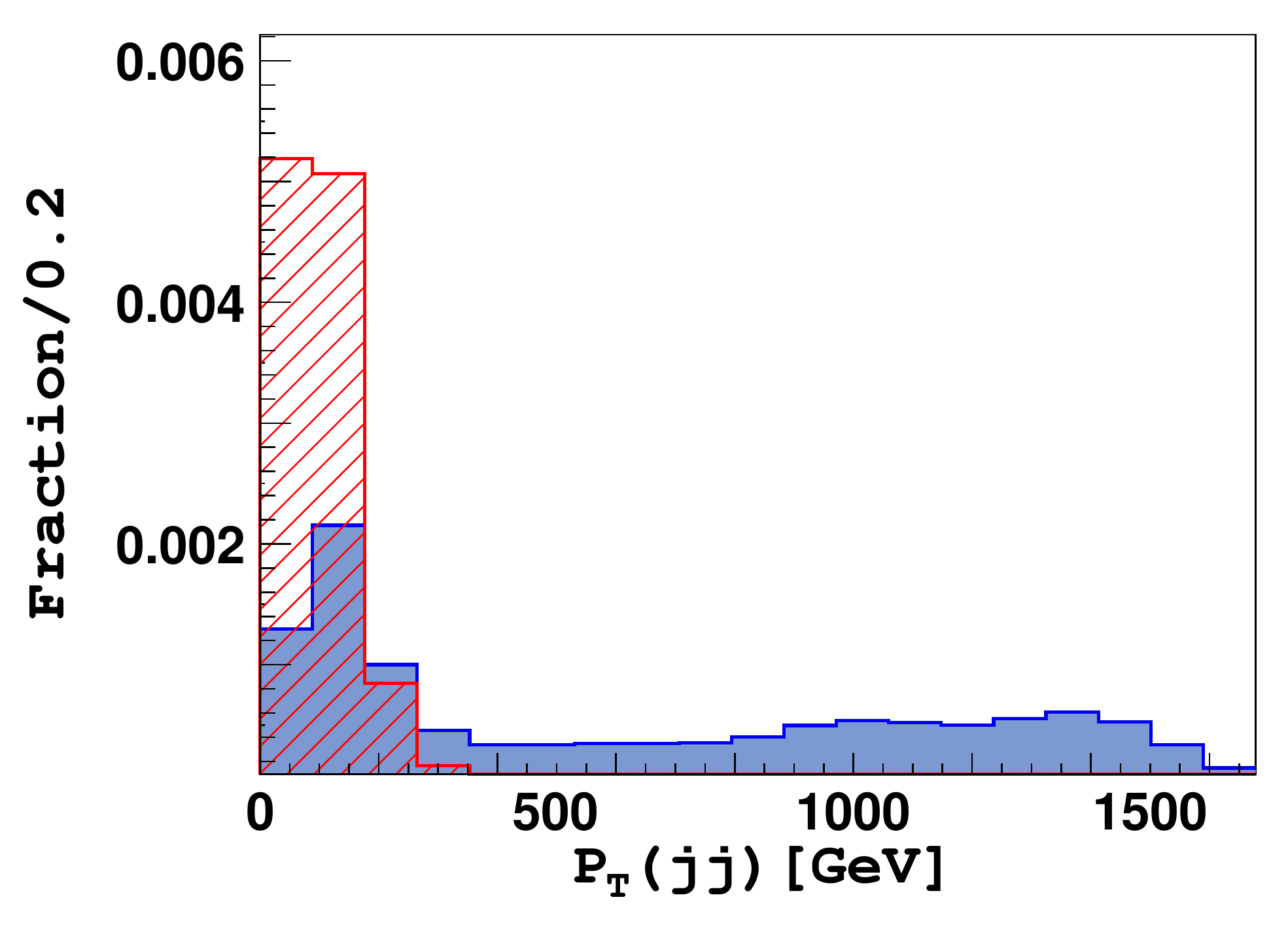}~~\includegraphics[width=5.cm,height=3.5cm]{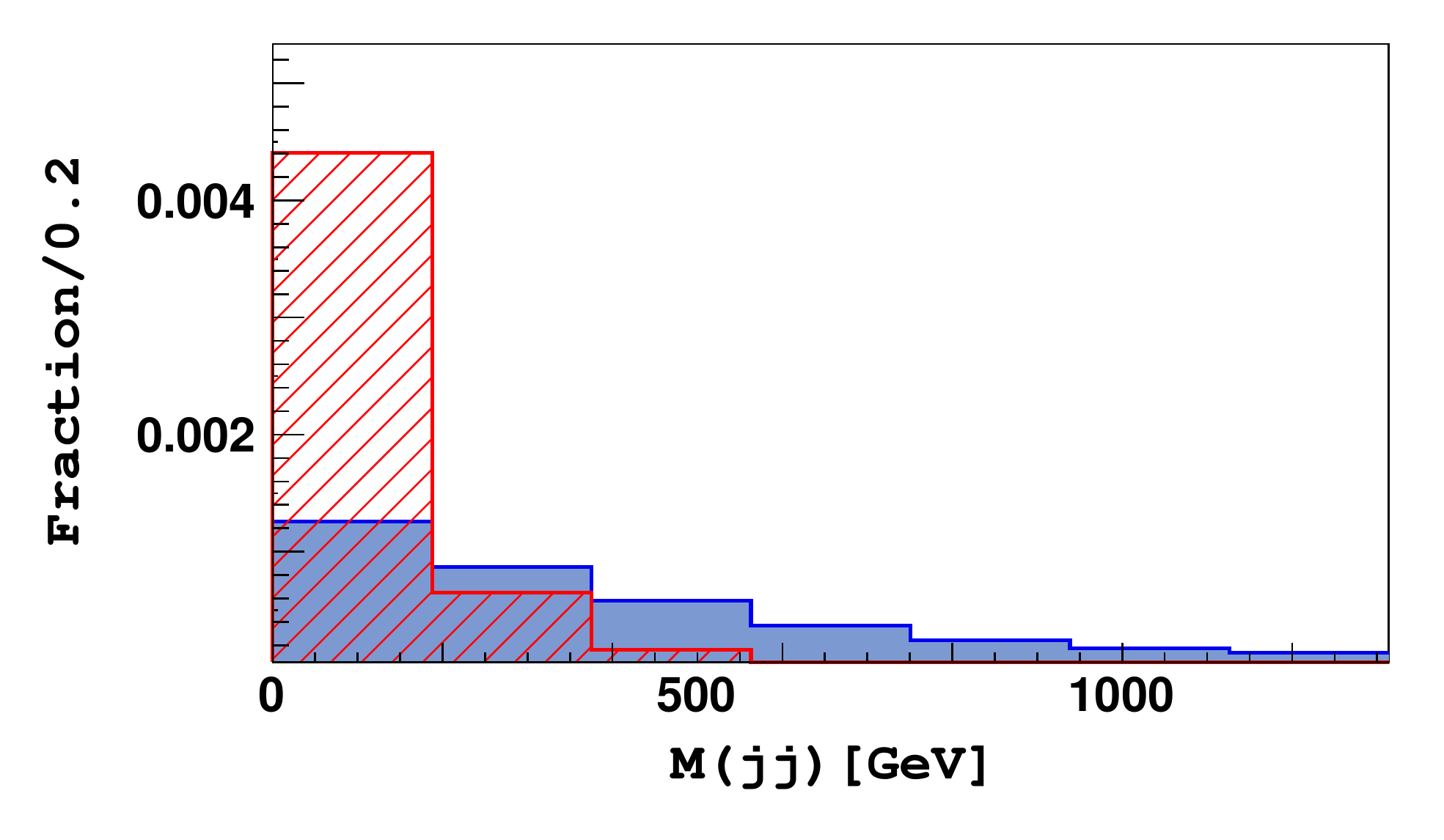}~~\includegraphics[width=5.cm,height=3.5cm]{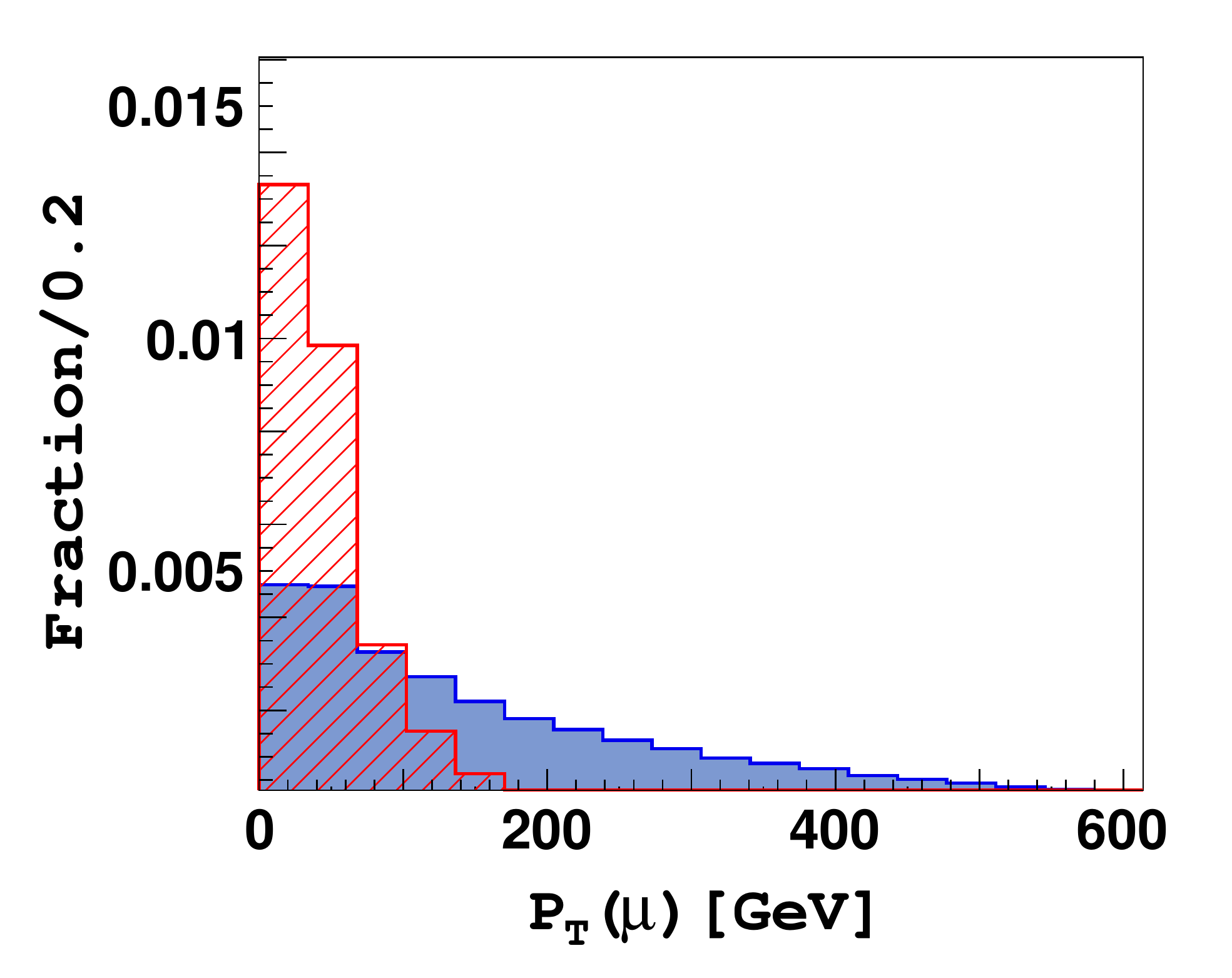}
\includegraphics[width=5.6cm,height=4cm]{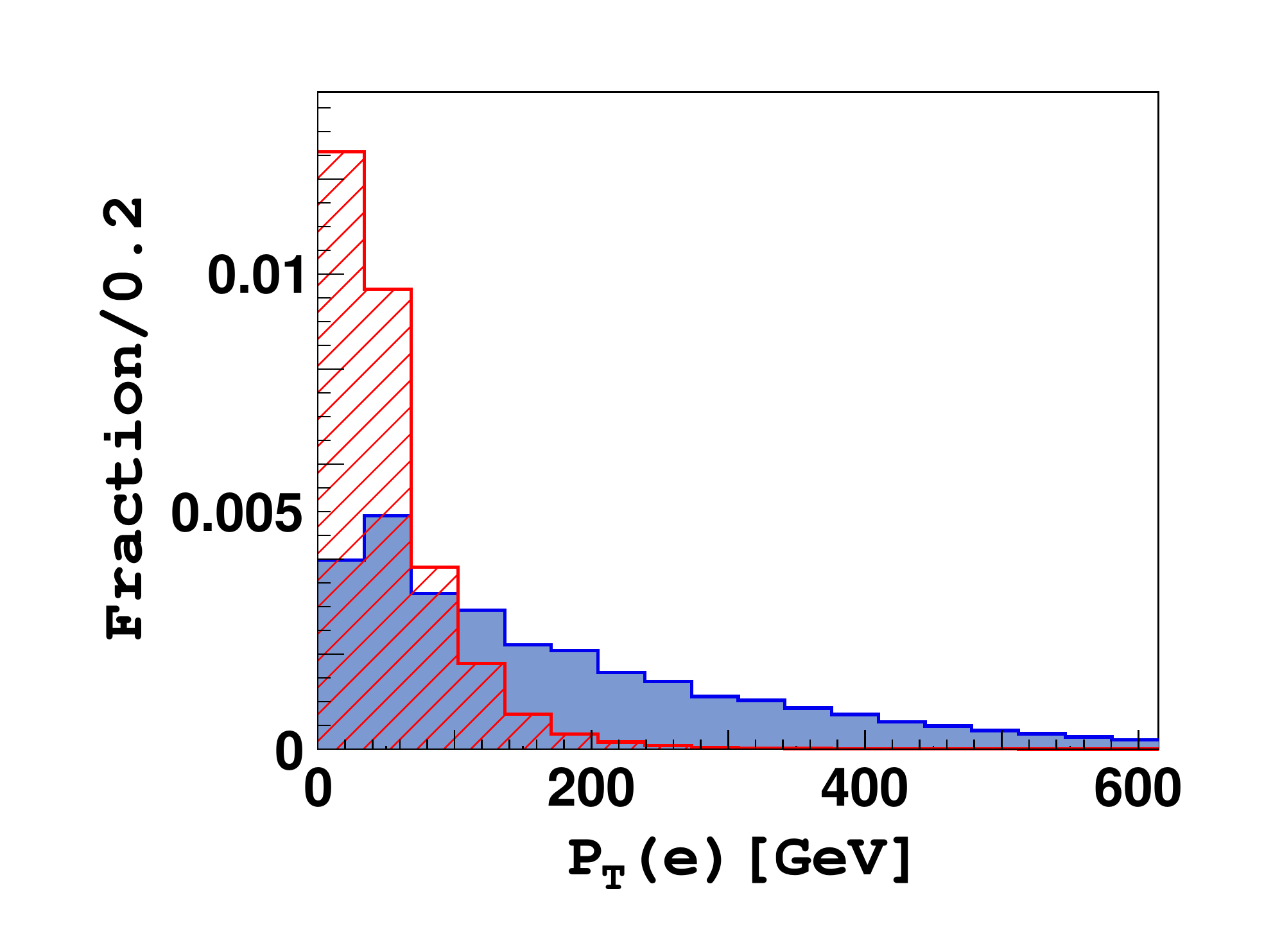}\includegraphics[width=5.cm,height=4cm]{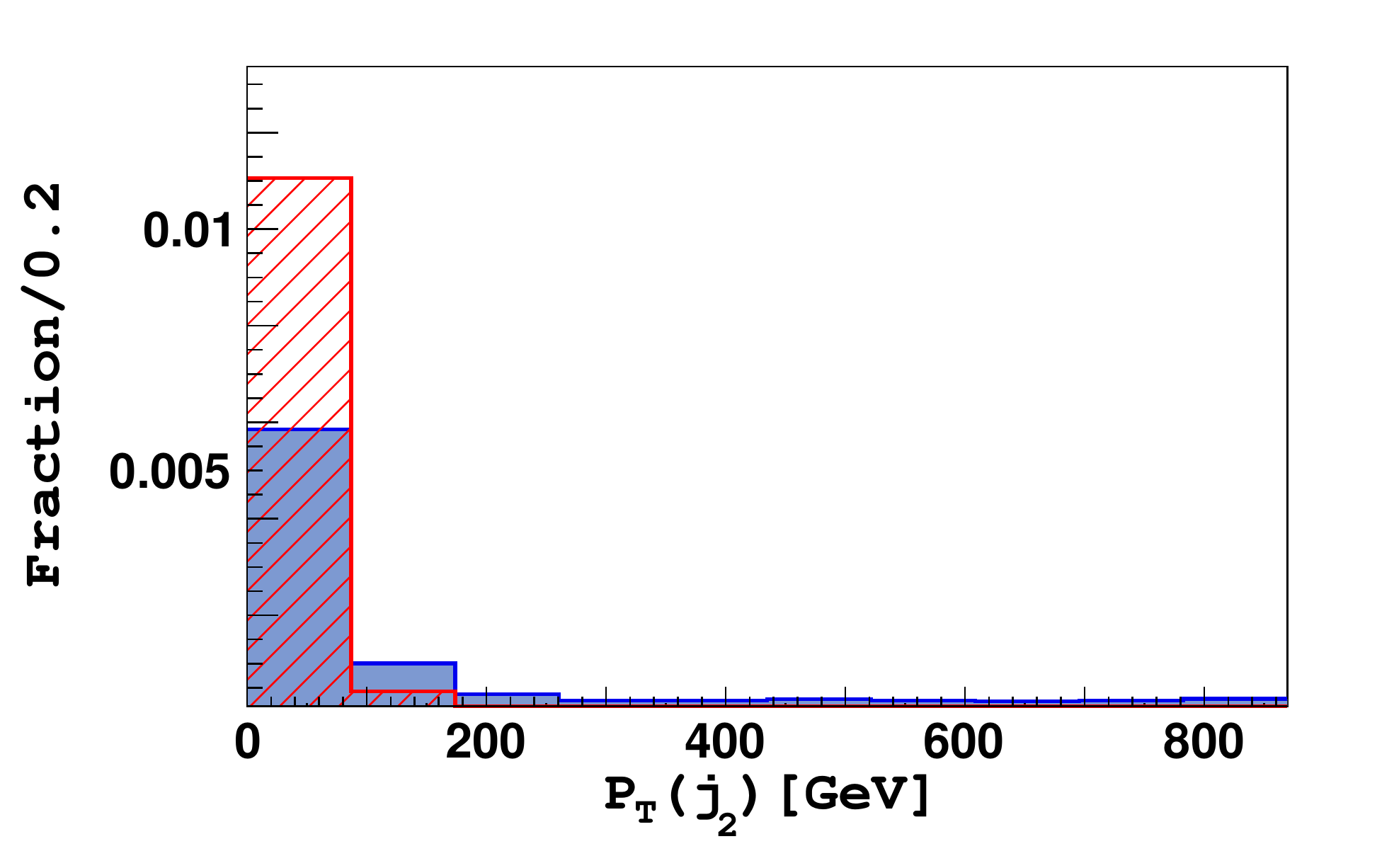}~~\includegraphics[width=5.cm,height=4cm]{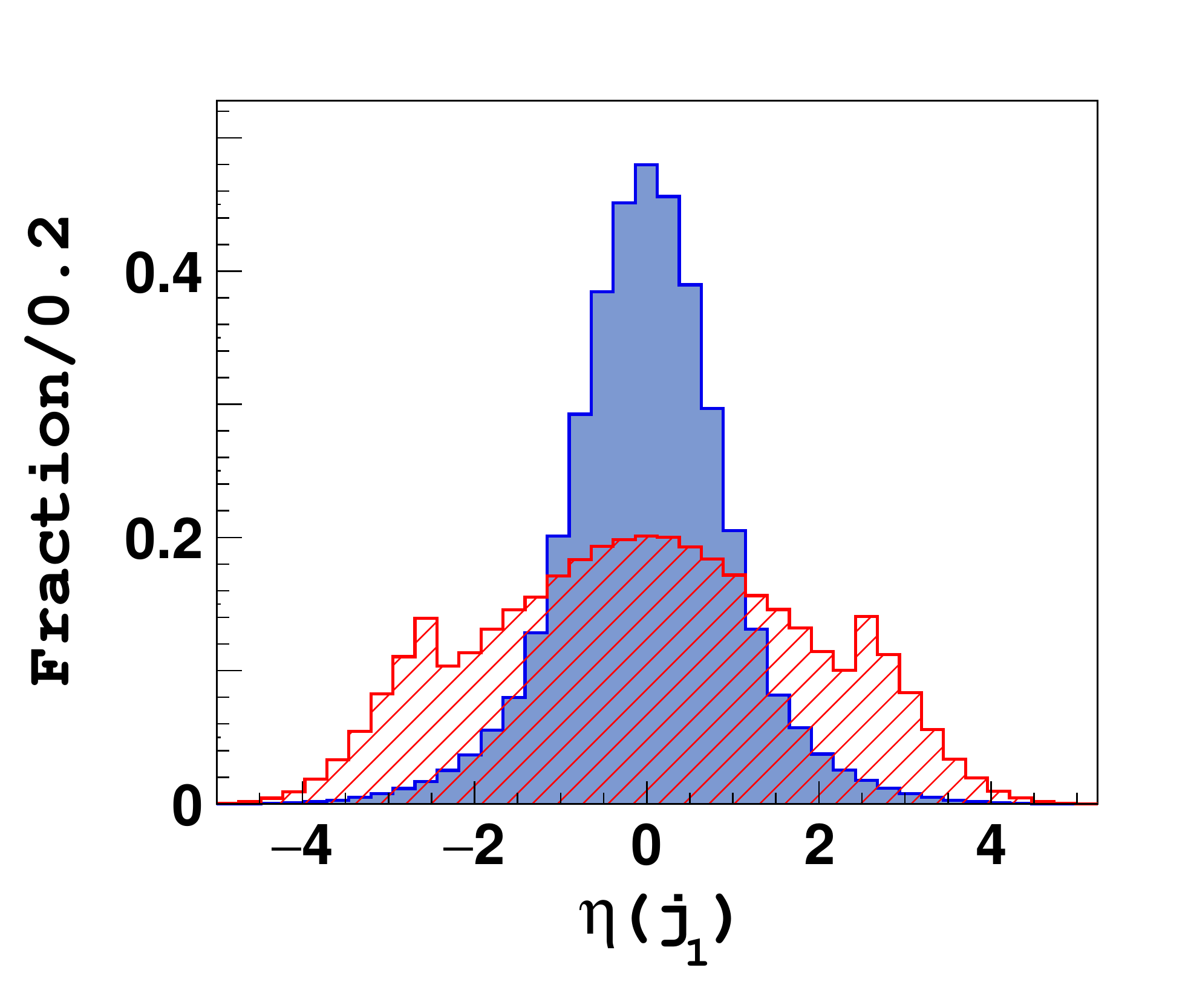}
\caption{{Input distributions to the BDT for signal events in blue and all relevant backgrounds in red.}}
\label{var}
\end{figure*}
Given the Feynman diagrams underpinning mono-$Z$ production in the BLSSM case for sneutrino  DM (see Fig.~\ref{fyn}{, top panels}), the $Z$ boson decaying leptonically can be reconstructed as such by constraining the emerging electron and muon pairs to reproduce $M_Z$ within experimental di-lepton mass resolution (we will not include $Z\to$ jet decays in the signal definition). 
The dominant irreducible background is $ZZ\to l^+ l^- \bar\nu\nu$  and the other large noise in this category is $W^+W^-\to l^+ \nu l^- \bar\nu $. As we reconstruct the $Z$ boson (specifically, by selecting the lepton pair that gives the closest value to the measured mass of the $Z$ boson),  the reducible backgrounds must contain $Z\to l^+l^-$. Given the hadronic environment of the LHC, additional jet activity is possible. Hence, the final list of backgrounds in this category is as follows: $Z+{\rm jets}$, $ZZ\to l^+ l^-+{\rm jets}$ and $ZW\to l^+ l^-  + {\rm jets}$. In addition, there are other reducible di-lepton backgrounds with jets  that we have dealt with: $t\bar{t} \to l^+ \nu b l^- \bar\nu\bar{b}$, as well as  $W^\pm +{\rm jets}$, which is reduced by an {{MET cut}}.  The last quantitatively important  background, purely leptonic,  is $ZW\to  l^+ l^- l\nu$, with one electron misidentified as jet. 
As preselection cuts we require $p_T(l) > 10$~GeV, $p_T(j) > 20$~GeV, $|\eta(l/j)| < 2.5 $ (where $j$ represents any jet and $l$ any lepton) and $\met > 50$~GeV. Tab.~\ref{tab:signatures4} shows the signal and background composition, wherein we emphasise the dominance of $S_2$ over $S_1$ owing to the
$\tilde\nu_i$ multiplicity in the former, while the latter only sees the involvement of $\tilde\nu_1$.  
Moreover, we stress that, while the signal is mediated by a heavy gauge boson, $Z'$, that leads to large MET,  the whole background is not,  thus we will eventually force the  $\met > 100$~GeV condition into the {BDT}. 
\begin{figure*}[t!]
\centering
\includegraphics[width=8.5cm,height=4.5cm]{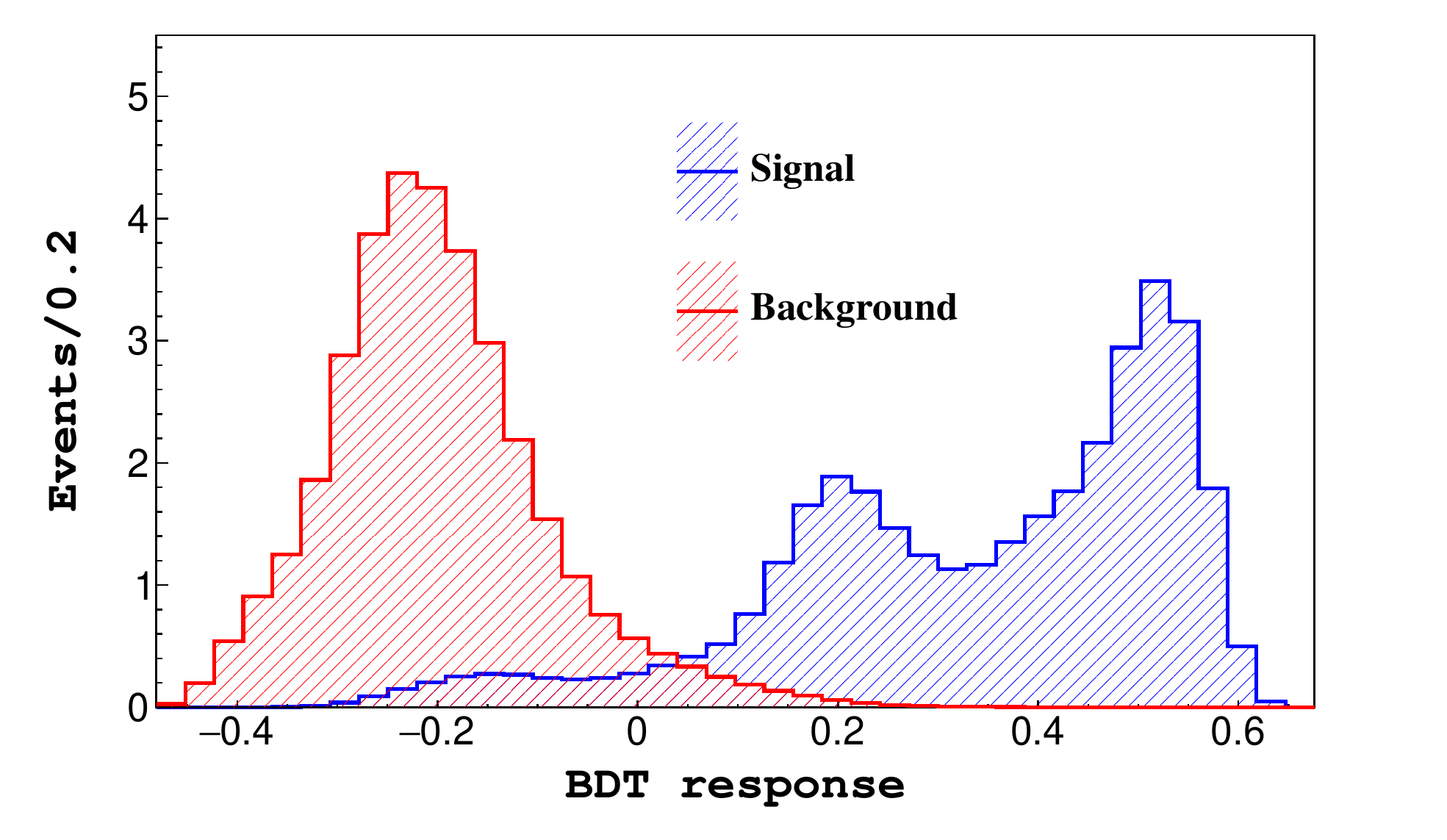}~
\includegraphics[width=8.5cm,height=4.5cm]{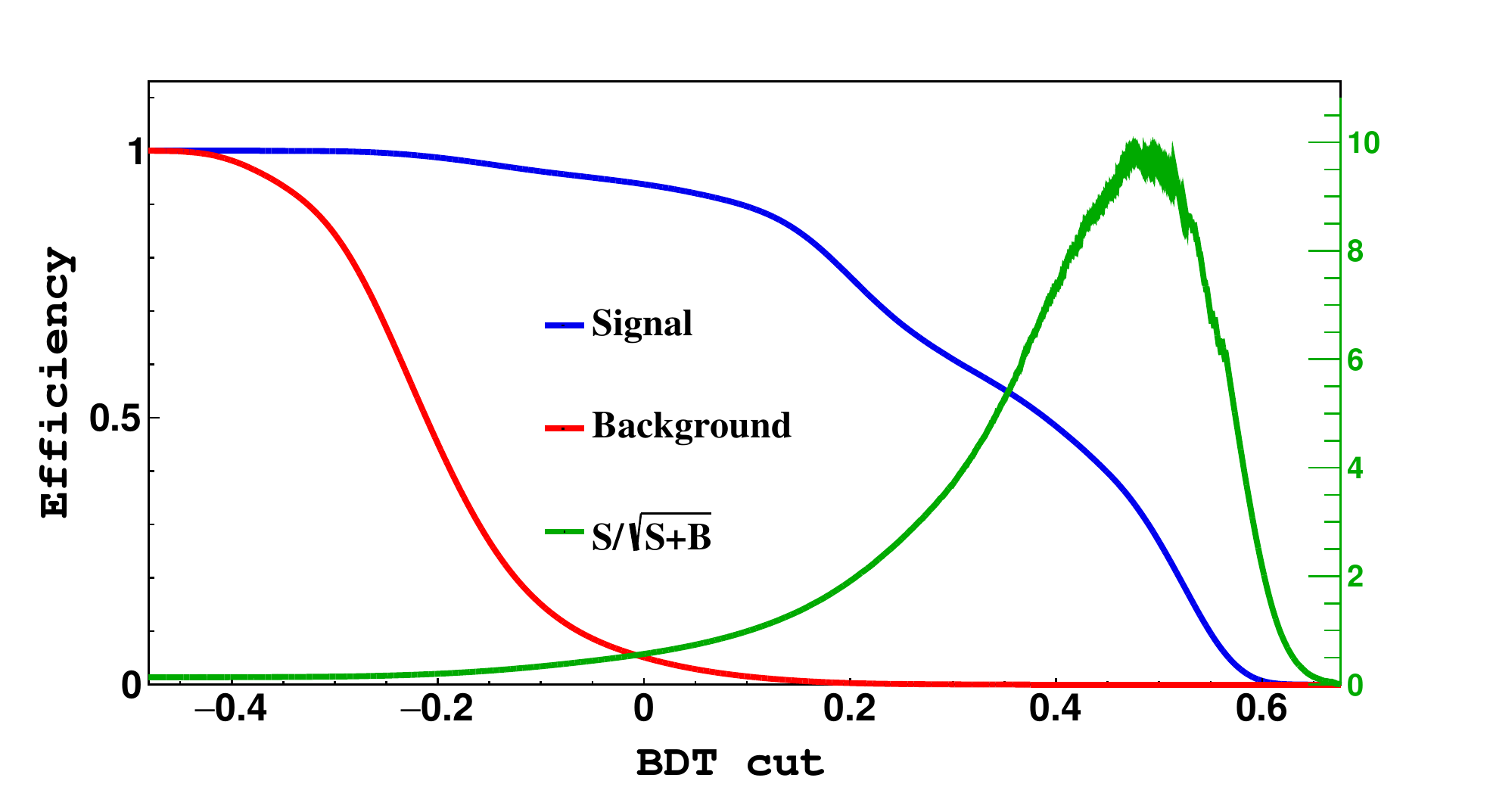}\caption{Left: BDT {response} for the signal in blue and background in red. Right: BDT cut efficiency for signal in blue and  background in red with the corresponding significance in green.}
\label{BDT-fig}
\end{figure*}
\begin{table}[h!]
\begin{center}
\begin{tabular}{|c|c|c|}
\hline
\multicolumn{2}{|c|}{Process} & $\sigma_{\rm tot} [{\rm pb}]$\Tstrut\Bstrut   \\ \hline\hline
${S_1}$&$pp\to Z^\prime Z (Z^\prime\to\tilde{\nu}_1\tilde{\nu}_1), (Z\to ll)$  &0.0041\Tstrut\Bstrut \\ \hline
${S_2}$&$pp\to Z^\prime\to\tilde{\nu}_i\tilde{\nu}_1 (\tilde{\nu}_i\to\tilde{\nu}_1 Z, Z\to ll)$ & 0.0115\Tstrut\Bstrut \\ \hline
\multirow{4}{*}{\rotatebox{90}{{Backgrounds}~~}}&$pp\to ZZ\to ll \nu \nu$&0.1256\Tstrut\Bstrut \\ \cline{2-3}
&$pp\to WW\to ll \nu \nu$ &1.013\Tstrut\Bstrut  \\ \cline{2-3}
&$pp\to ZW\to lll \nu$ &0.129\Tstrut\Bstrut \\ \cline{2-3}
&$pp\to W j \to l\nu +j$  &2008\Tstrut\Bstrut \\\cline{2-3}
&$pp\to  t\bar{t} $&597\Tstrut\Bstrut \\ \hline
\end{tabular}
\end{center}
\caption{Total cross section in pb for the signal (split into the two topologies of Fig.~\ref{fyn}{, top panels})  and the dominant background processes considered in our analysis. The samples have been produced with the following cuts: $p_T(l)> 10$~GeV, $p_T(j)> 20$~GeV and $\met > 50$~GeV. }
\label{tab:signatures4}
\end{table}

Upon enforcing all kinematic conditions above, 
relevant  distributions are given as an input to our BDT in order  to perform a Multi-Variate Analysis (MVA)~\cite{TMVA2007}. {{An important feature of the MVA is that it can rank the input variables according to their ability to separate between signal   and  background  events. For illustrative purposes, we show the first two variables ranked by the BDT for the signals $S_1$ and $S_2$ as well as all  backgrounds separately in Fig.~\ref{Mll}. Herein,  a peculiar feature is the fact that the signal is mediated by a heavy $Z^\prime$, so that this causes the ISR jet in $S_2$  to recoil against a very massive object. Such kinematics pushes the transverse momentum distribution of the leading jet to be peaked around half of $Z^\prime$ mass. This does not occur for $S_1$, though, owing to the presence of also the ISR $Z$. For the di-lepton transverse momentum, both signals have a much stiffer spectrum than any of the backgrounds, again, owing to $Z$ balancing the heavy $Z'$ (in $S_1$) or else being ejected by the decay of the latter at large $p_T$ (in $S_2$).}}

The discriminating power of the BDT relies on the fact that the signals and  backgrounds may be characterised by different features that can be encoded into several distributions.   For completeness, we sketch the first 9 most important variables, as ranked by the BDT,  in Fig.~\ref{var} (wherein backgrounds are shown cumulatively). Further, Tab.~\ref{tab:rank}  shows the BDT ranking of all input variables according to the their power in separating the signal and  background events. Our ML approach is then  based on a set of BDTs where each tree yields a binary output depending on whether an event is classified as signal- or background-like during the training session. The most important feature of the MVA algorithm is its possibility to combine the various discriminating kinematic distributions into one main discriminator, the BDT response, and thus dealing with only one variable to maximise the signal  rate over the background one. The BDT response ranges between $-1$ and $+1$ corresponding to pure background and pure signal, respectively.

\begin{table}[h!]
\begin{center}

\begin{tabular}{|c|c|c|}
  \hline
  Rank & Variable & Separation power \\
  \hline
  \hline 
  1 & $p_T(j_1)$  & $61.79\%$  \\
 \hline 
  2 & $p_T(ll)$  & $52.09\%$  \\  
  \hline 
  3 & $\Delta R(l,l)$  & $48.35\%$  \\  
  \hline 
  4 & $p_T(jj)$  & $43.81\%$  \\  
  \hline 
  5 & $M(jj)$  & $40.90\%$  \\  
  \hline 
  6 & $p_T(\mu)$  & $29.69\%$  \\  
  \hline 
  7 & $p_T(e)$  & $29.66\%$  \\  
  \hline 
  8 & $p_T(j_2)$  & $28.00\%$  \\  
  \hline 
  9 & $\eta(j_1)$  & $18.44\%$  \\  
  \hline 
  10 & $\eta(jj)$  & $8.173\%$  \\  
  \hline 
  11 & $\Delta R(jj,l^-)$  & $6.434\%$  \\  
   \hline 
  12 & $\Delta R(jj,l^+)$  & $6.129\%$  \\  
   \hline 
  13 & $\eta(ll)$  & $6.112\%$  \\  
   \hline 
  14 & $\eta(j_2)$  & $5.281\%$  \\  
   \hline 
  15 & $\eta(\mu)$  & $5.189\%$  \\  
   \hline 
  16 & $\Delta R(ll,j_1)$  & $5.002\%$  \\  
   \hline 
  17 & $\met$  & $4.928\%$  \\  
   \hline 
  18 & $\Delta R(j_1,j_2)$  & $4.658\%$  \\  
   \hline 
  19 & $\eta(e)$  & $ 4.571\%$  \\  
   \hline 
  20 & $\Delta R(ll,j_2)$  & $4.358\%$  \\  
   \hline 
  21 & $M(ll)$  & $2.713\%$  \\  
    \hline
\end{tabular}
\end{center}
\caption{{BDT ranking  of the input variables in descending order of  discriminative power.} }
\vspace{-0.2cm}
\label{tab:rank}
\end{table}

After the aforementioned kinematic cuts (preselection), the {total} number of events for the signal is $656$ while for the background is  ${2.3}\times 10^{7}$, both of which are passed to the MVA environment to perform the ML analysis. The resulting  BDT {response} is shown in Fig.~\ref{BDT-fig}~(left) with signal events in blue and background ones  in red. Enhancing the BDT cut efficiency is done by maximising the function $S/\sqrt{S+B}$, where $S$ is the total signal rate and $B$ {is} the background one at the given luminosity. Hence, for the optimal value of the BDT cut set at $0.48$, the remaining signal events  (222) and  background ones (285) yield a significance of {$9.8\sigma$}. This corresponds to a signal extraction efficiency of {$34\%$} and a background rejection efficiency of $1.2\times 10^{-5}$. Fig.~\ref{BDT-fig} (right) shows the signal efficiency in blue and the background rejection efficiency  in red versus the BDT cut with the corresponding significance in green.

Finally, notice that the analysis has been performed at a Center-of-Mass (CM) energy of $14$~TeV and integrated luminosity of 100~fb$^{-1}$. For the simulation of the signal and background event samples, we have used  {MadGraph5 
({v2.4.3})~\cite{Alwall:2014hca}. Parton shower and hadronisation have been carried out by {PYTHIA6}~\cite{Sjostrand:2006za, Sjostrand:2007gs} while a fast detector simulation by {Delphes}~\cite{deFavereau:2013fsa} was used.


\section{Conclusions}
\label{sec.5}
We have shown that a ML based approach, as opposed to a standard cut-flow one, is well suited to extract a mono-$Z(\to l^+l^-)$ signal 
of the BLSSM at the LHC, with
14~TeV and 100~fb$^{-1}$ of energy and luminosity, respectively. The latter is emerging from a heavy $Z'$ boson decaying into sneutrinos, the lightest of which is the DM state  of this scenario, eventually yielding a di-lepton plus MET signature with additional jet activity. Furthermore, the ability of the $Z$ boson to couple directly to the DM state enables one to access the spin properties of the latter, specifically, by studying the angular behaviour of either lepton relative to the $Z$ boson direction in its rest frame. We have illustrated this phenomenology using a single benchmark point in the BLSSM, compliant with current experimental limits. We defer to a future publication the illustration of such an approach applied to the entire BLSSM parameter space {\cite{wip}}.

\section*{Acknowledgments}
W.A. acknowledges the support from the Department of Atomic Energy (DAE) `Neutrino Project' of the Harish-Chandra Research Institute (HRI). The work of A.H. is supported by
the Swiss National Science Foundation. The work of S.K. and S.M. was partially supported by the H2020-MSCA-RISE-2014 grant No.~645722 (NonMinimalHiggs). S.M. is financed in part through the NExT Institute. S.K. acknowledges partial support from the  Durham IPPP Visiting Academics (DIVA) programme.



\begin{thebibliography}{99}

\bibitem{Khalil:2015naa}
  S.~Khalil and S.~Moretti,
  Rept.\ Prog.\ Phys.\  {\bf 80} (2017) no.~3,  036201
  [arXiv:1503.08162 [hep-ph]].

\bibitem{Abdallah:2017gde}
  W.~Abdallah and S.~Khalil,
  JCAP {\bf 1704} (2017) no.~4,  016
  [arXiv:1701.04436 [hep-ph]].

\bibitem{DelleRose:2017uas}
  L.~Delle Rose, S.~Khalil, S.~J.~D.~King, S.~Kulkarni, C.~Marzo, S.~Moretti and C.~S.~Un,
  JHEP {\bf 1807} (2018) 100
  [arXiv:1712.05232 [hep-ph]].

\bibitem{DelleRose:2017ukx}
  L.~Delle Rose, S.~Khalil, S.~J.~D.~King, C.~Marzo, S.~Moretti and C.~S.~Un,
  Phys.\ Rev.\ D {\bf 96} (2017) no.~5,  055004
  [arXiv:1702.01808 [hep-ph]].

\bibitem{Abdallah:2015uba}
  W.~Abdallah, J.~Fiaschi, S.~Khalil and S.~Moretti,
  JHEP {\bf 1602} (2016) 157
  [arXiv:1510.06475 [hep-ph]].

\bibitem{Abdallah:2016vcn}
  W.~Abdallah, A.~Hammad, S.~Khalil and S.~Moretti,
  Phys.\ Rev.\ D {\bf 95} (2017) no.~5,  055019
  [arXiv:1608.07500 [hep-ph]].

\bibitem{Khalil:2016lgy}
  S.~Khalil,
  Phys.\ Rev.\ D {\bf 94} (2016) no.~7,  075003
  [arXiv:1606.09292 [hep-ph]].

\bibitem{Khalil:2010iu}
  S.~Khalil,
  Phys.\ Rev.\ D {\bf 82} (2010) 077702
  [arXiv:1004.0013 [hep-ph]].
        

\bibitem{Bechtle:2008jh}
  P.~Bechtle, O.~Brein, S.~Heinemeyer, G.~Weiglein and K.~E.~Williams,
  Comput.\ Phys.\ Commun.\  {\bf 181} (2010) 138
  [arXiv:0811.4169 [hep-ph]].
  
\bibitem{Bechtle:2013xfa}
  P.~Bechtle, S.~Heinemeyer, O.~St{\aa}l, T.~Stefaniak and G.~Weiglein,
  Eur.\ Phys.\ J.\ C {\bf 74} (2014) no.~2,  2711
  [arXiv:1305.1933 [hep-ph]].

\bibitem{Alves:2015dya} 
  A.~Alves and K.~Sinha,
  Phys.\ Rev.\ D {\bf 92}, no. 11, 115013 (2015)
  [arXiv:1507.08294 [hep-ph]].
  
\bibitem{Alves:2017uls} 
  A.~Alves, A.~C.~O.~Santos and K.~Sinha,
  Phys.\ Rev.\ D {\bf 97}, no. 5, 055023 (2018)
  [arXiv:1710.11290 [hep-ph]].

\bibitem{Bernreuther:2018nat} 
  E.~Bernreuther, J.~Horak, T.~Plehn and A.~Butter,
  SciPost Phys.\  {\bf 5}, 034 (2018)
  [arXiv:1805.11637 [hep-ph]].


\bibitem{Aaboud:2018sfi} 
  M.~Aaboud {\it et al.} [ATLAS Collaboration],
  Phys.\ Lett.\ B {\bf 793}, 499 (2019)
  [arXiv:1809.06682 [hep-ex]].
  
\bibitem{Sirunyan:2018owy} 
  A.~M.~Sirunyan {\it et al.} [CMS Collaboration],
  Phys.\ Lett.\ B {\bf 793}, 520 (2019)
  [arXiv:1809.05937 [hep-ex]].

  
\bibitem{Choi:2003fs}
  S.~Y.~Choi and Y.~G.~Kim,
  Phys.\ Rev.\ D {\bf 69} (2004) 015011
  [hep-ph/0311037].
  
  
  
\bibitem{Choi:2018sqc}
  S.~Y.~Choi,
  Phys.\ Rev.\ D {\bf 98} (2018) no.~11,  115037
  [arXiv:1811.10377 [hep-ph]].
  
  
\bibitem{Bagger:1996bt}
  J.~A.~Bagger, K.~T.~Matchev, D.~M.~Pierce and R.~j.~Zhang,
  Phys.\ Rev.\ D {\bf 55} (1997) 3188
  [hep-ph/9609444].
  
\bibitem{Ghosal:1997dv}
  A.~Ghosal, A.~Kundu and B.~Mukhopadhyaya,
  Phys.\ Rev.\ D {\bf 57} (1998) 1972
  [hep-ph/9709431].

\bibitem{Dutta:2019gox}
  J.~Dutta, B.~Mukhopadhyaya and S.~K.~Rai,
  arXiv:1904.08906 [hep-ph].

  
\bibitem{TMVA2007}
        A.~Hoecker, P.~Speckmayer, J.~Stelzer,
        J.~Therhaag, E.~von Toerne, and H.~Voss,
       CERN-OPEN-2007-007  
[physics/0703039]. 
  
  
\bibitem{Alwall:2014hca} 
  J.~Alwall {\it et al.},
  JHEP {\bf 1407}, 079 (2014)
  [arXiv:1405.0301 [hep-ph]].
  
\bibitem{Sjostrand:2006za} 
  T.~Sjostrand, S.~Mrenna and P.~Z.~Skands,
  JHEP {\bf 0605}, 026 (2006)
  [hep-ph/0603175].
\bibitem{Sjostrand:2007gs} 
  T.~Sjostrand, S.~Mrenna and P.~Z.~Skands,
  Comput.\ Phys.\ Commun.\  {\bf 178}, 852 (2008)
  [arXiv:0710.3820 [hep-ph]].
\bibitem{deFavereau:2013fsa} 
  J.~de Favereau {\it et al.} [DELPHES 3 Collaboration],
  JHEP {\bf 1402}, 057 (2014)
  [arXiv:1307.6346 [hep-ex]].
  %
\bibitem{wip}
W.~Abdallah, A.~Hammad, S.~Khalil and S.~Moretti, work in progress.
%
\end{thebibliography}
\end{document}